\documentclass[twocolumn,prl,a4paper,showpacs,floatfix,preprintnumbers,amsmath,amssymb]{revtex4}

\usepackage{graphicx}
\usepackage{dcolumn}

\newcommand{\ud}{\mathrm{d}}

\begin{document}

\title{Rabi Interferometry and Sensitive Measurement \\
of the Casimir-Polder Force with Ultra-Cold Gases}

\author{Jan Chwede\'nczuk$^1$, Luca Pezz\'e$^2$, Francesco Piazza$^1$ and Augusto Smerzi$^1$}

\affiliation{$^1$BEC-CNR-INFM and Dipartimento di Fisica, Universit\`a di Trento, I-38050 Povo, Italy,\\
  $^2$Laboratoire Charles Fabry, Institut d'Optique, 2 Avenue Fresnel, 91127 Palaiseau - France}

\begin{abstract} 

  We show that Rabi oscillations of a degenerate fermionic or bosonic gas  
  trapped in a double-well potential can be exploited for the interferometric  
  measurement of external forces at micrometer length scales.
  The Rabi interferometer is less sensitive, but easier to implement, than the
  Mach-Zehnder since it 
  does not require dynamical beam-splitting/recombination processes. 
  As an application we propose a measurement of the 
  Casimir-Polder force acting between the atoms and a dielectric surface.
  We find that even if the interferometer is
  fed with a coherent state of relatively small number of atoms, and in the presence of realistic 
  experimental noise, the force can be measured with a sensitivity 
  sufficient to discriminate between thermal and zero-temperature regimes of the Casimir-Polder potential.
  Higher sensitivities can be reached with spin squeezed states.
\end{abstract}

\date{\today}

\maketitle

{\it Introduction}.
Interferometers with trapped ultra-cold atoms are valuable tools for the
precise measurement of external forces \cite{Cronin_2009}. 
A promising one is the double-well Mach-Zehnder interferometer (MZI)
\cite{Pezze_2005,Lee_2006,Huang_2008,Jo_2007,Schumm_2005,Gati_2006,Bohi_2009}.
This requires two 50/50 beam splitters implemented by a dynamical manipulation of the inter-well 
barrier. The phase shift is accumulated during the interaction of atoms with an external potential in the
absence of the inter-well coupling. 
It is interesting to search for alternative interferometric schemes
which can be easier to realize and therefore can be more stable than the MZI.
In this Letter we propose a new protocol to create a Rabi interferometer. 
It can be implemented using either degenerate spin-polarized Fermions or non-interacting 
Bose-Einstein condensates (BECs) trapped in a double-well potential. 
The gas tunnels between the two wells 
while acquiring, \emph{at the same time}, a phase shift.
The relative number of particles among the two wells undergoes Rabi oscillations 
analogous to those experienced by a collection of two-level atoms in a radio frequency field \cite{Windpassinger_2008}. 
The measurement of population imbalance as a function of time allows to infer the value of the external force as
it affects both the amplitude and frequency of Rabi oscillations. 
The Rabi interferometer is less sensitive than the MZI, 
but does not require any splitting/recombination processes
and is suitable for the estimation of forces rapidly decaying with distance.
In particular, once fed with a fermionic/bosonic spin coherent state, 
the interferometer allows for the accurate measurement of the Casimir-Polder force between the atomic sample and a surface. 
We show that even
in the presence of typical experimental noise it is possible to distinguish between 
thermal and zero-temperature regimes of the Casimir-Polder potential
\cite{Antezza_2004}, which has not yet been achieved in experiment
\cite{Pasquini_2004, Harber2005, Lin_2004, Obrecht_2007, sukenik_1993}. 
Moreover, we demonstrate that the Rabi interferometer can further benefit from the use of entangled states as input. 
In analogy to the MZI, a sub shot-noise phase sensitivity can be obtained with spin squeezed states 
recently created with a BEC \cite{Esteve_2008}.

{\it The Rabi interferometer}.
Let us consider a degenerate gas of non-interacting atoms confined in a double-well potential along the $x_1$
direction, and in a harmonic trap along $x_2$ and $x_3$.
A field operator formalism
allows for studying interferometry with both bosons and fermions.
We introduce the field operator 
$\hat{\Psi}(\vec{r}) = \sum_{\vec{k}} \psi_{k_1}(x_1)\psi_{k_2}(x_2)\psi_{k_3}(x_3) \, \hat{c}_{\vec{k}}$, 
where $\psi_{k_i}(x_i)$ are single-particle energy eigenfunctions along the $i$-th direction
and the sum is over the complete set of quantum numbers $\vec{k}\equiv(k_1,k_2,k_3)$.
The annihilation (creation) operators $\hat{c}_{\vec{k}}$ ($\hat{c}_{\vec{k}}^\dag$) 
satisfy commutation or anticommutation relations, depending on the statistics.
In two-mode approximation (taking into account only lowest $k_1=0$ and the first excited $k_1=1$ states)
we introduce the wave functions
$\psi_{a/b}(x_1) = (\psi_{0}(x_1) \pm \psi_{1}(x_1))/\sqrt{2}$
and mode operators $\hat{a}_{\vec{k}_\perp}= (\hat{c}_{0,\vec{k}_\perp}+\hat{c}_{1,\vec{k}_\perp})/\sqrt{2}$
and $\hat{b}_{\vec{k}_\perp} = (\hat{c}_{0,\vec{k}_\perp}-\hat{c}_{1,\vec{k}_\perp})/\sqrt{2}$,
where $\vec{k}_\perp\equiv (k_2,k_3)$. The dynamics of the system is thus governed by the Hamiltonian
\begin{equation}\label{ham}
  \hat{H}=-E_J \hat{J}_x+\delta \hat{J}_z,
\end{equation}
where $E_J$ is the tunneling energy, $\delta$ is the relative energy shift due to interaction
with a position-dependent external potential $V(x_1)$ \cite{nota_integrals}. The
operators $\hat J_x$, $\hat J_y$ and $\hat J_z$ \cite{nota_J}
form a closed algebra of angular momentum.
The goal of the Rabi interferometer is to estimate $\delta$ with the highest possible sensitivity.
We consider the measurement of the population imbalance between the two modes,
which can be expressed in terms of the eigenvalues of the operator $\hat{J}_z$. 
Using the evolution operator generated by (\ref{ham}),
\begin{equation}\label{evo_new}
\hat U(t)=e^{-i\alpha\hat J_y}e^{i \Omega\hat J_x}e^{i\alpha\hat J_y},
\end{equation}
we obtain $\hat J_z(t, \delta)=u(t, \delta) \hat J_x+v(t, \delta) \hat J_y+w(t, \delta) \hat J_z$,
with $u(t, \delta)=\sin\alpha\cos\alpha(\cos\Omega-1)$, $v(t, \delta)=-\cos\alpha\sin\Omega$ and
$w(t, \delta)=(\cos^2\alpha\cos\Omega+\sin^2\alpha)$. 
Here, $\Omega=\omega t$, $\cos\alpha=E_J/\hbar\omega$, $\sin\alpha=\delta/\hbar\omega$ and
$\omega=\sqrt{E_J^2+\delta^2}/\hbar$ is the detuned Rabi frequency.

The estimation protocol consists of measuring the population imbalance at $k$ times $t_1,t_2,..., t_k$,
with $m$ repetitions at each time.
The value of $\delta$ is estimated by $\delta_{est}$ resulting from a least squares fit of a 
theoretical curve $\langle \hat J_z(t, \delta_{\mathrm{est}}) \rangle$
to the set $\{n\}=(n_1,\ldots,n_k)$ of acquired points.
In order to determine the error on $\delta_{est}$, we notice that if $m\gg1$, 
according to the central limit theorem, the conditional probability 
for measuring a value $n_i$ of the population imbalance at time $t_i$ tends to
$  p(n_i |\delta)=\frac{1}{\sqrt{2\pi} \Delta \hat{J}_z(t_i,\delta) /\sqrt m}
\exp \big[-\frac{(n_i- \langle \hat J_z(t_i,\delta) \rangle )^2}{2 \Delta^2\hat{J}_z(t_i,\delta)/m} \big]$, 
where $\Delta^2\hat{J}_z(t_i,\delta) = \langle \hat{J}_z(t_i,\delta)^2 \rangle - \langle \hat{J}_z(t_i,\delta) \rangle^2$.
The average values are calculated on the input state $|\psi_{inp}\rangle$.
Since measurements at different times are independent, 
the joint conditional probability of detecting $\{n\}$ reads
$  p(\{n\}|\delta)=\prod_{i=1}^kp(n_i|\delta)$.
It is then possible to demonstrate that $\delta_{\mathrm{est}}$
corresponds to $\delta_{\mathrm{ML}}$, called the maximum likelihood (ML),  
which maximizes the probability $p(\{n\}|\delta)$ \cite{nota_ML}. 
Finally, the error on $\delta_{\mathrm{ML}}$, and thus on $\delta_{\mathrm{est}}$, is given
by the inverse of the Fisher information, $\Delta^2\delta_{\rm ML}=\mathcal{F}^{-1}$ \cite{nota_Fisher}
and reads $\Delta^2\delta_{\rm ML}=\big(\sum_{i=1}^k \frac{1}{\Delta^2\delta(t_i)} \big)^{-1}$, where 
\begin{equation}\label{err}
  \Delta^2\delta(t) =\frac{\Delta^2\hat{J}_z(t,\delta)}
        {m\left[\frac{\partial}{\partial\delta}\langle \hat{J}_z(t,\delta) \rangle\right]^2}.
\end{equation}
Let us now consider a coherent spin state (CSS) \cite{Arecchi_1972} as input of the Rabi interferometer.
This state corresponds to a Poissonian distribution of particles among the two wells. 
For fermions, it is given by
$|\mathrm{CSS}\rangle_F = \prod_{ \vec{k}_{\perp}} \hat{c}^\dag_{0,\vec{k}_{\perp}} |0\rangle$, 
where $|0\rangle$ is the vacuum and the product runs over the first $N$ excited states along the 
($x_2,x_3$) directions, while
for bosons $|\mathrm{CSS}\rangle_B = N!^{-1/2}(\hat{c}^\dag_{0,\vec{k}_{\perp}=0})^N |0\rangle$.
When the interferometer is fed with the CSS, the relative population oscillates as
\begin{equation}
  \langle \hat J_z(t, \delta) \rangle = \frac{N}{2} \tan\alpha \big( \cos\Omega -1 \big),
\end{equation}
and the sensitivity 
\begin{eqnarray} \label{Ddelta}
  \frac{\Delta \delta(t)}{\delta} 
  = \frac{1}{\sqrt m\sqrt{N}\tan\alpha}  \bigg| \cos\Omega-1- \frac{\sin^2\alpha}{\cos\alpha}\Omega \sin\Omega \bigg|^{-1}, 
\end{eqnarray}
scales at the shot noise
limit, $\frac{\Delta \delta(t)}{\delta} \sim 1/\sqrt{N}$.
The sensitivity of the Rabi interferometer fed by a CSS scales as $t^{-1}$ only when 
$\frac{\sin^2\alpha}{\cos\alpha}\Omega\geq1$. Using realistic values for 
$\delta$ and $E_J$ (see following section), the above condition is satisfied for $t\gtrsim3$ s. This implies,
that when the measurements are done within first few Rabi periods, the sensitivity does not benefit from
the scaling with time. This is in contrast to the MZI, where $\Delta\delta\propto t^{-1}$ even for short times.

\begin{figure}[htb!]
\includegraphics[clip, scale=0.35]{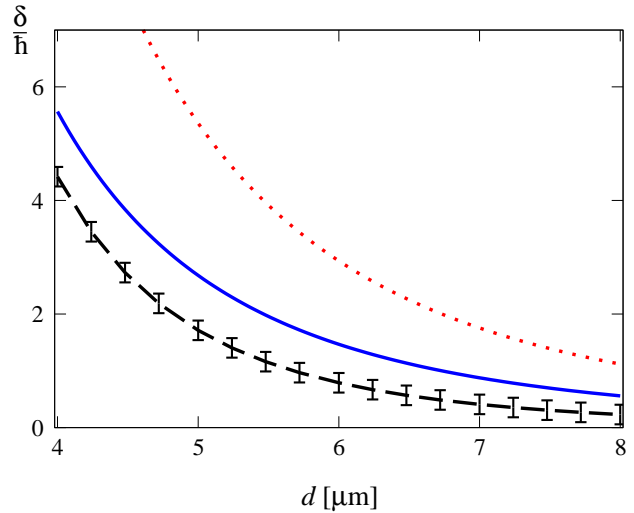}
\caption{(color online): The detuning $\frac{\delta}{\hbar}$ calculated with $V_\mathrm{CP}(x_1;d)$
  (black dashed line)
  and the corresponding sensitivity (error bars) of
  a fit to $k=10$ points in time and $m=10$ measurements at each time point. The uncertainty
  includes the effect of atom-atom interactions and limited resolution of the measurement of population imbalance
  (see text for details). 
  The input state is the classical spin coherent state. 
  Also, a detuning calculated with $V^{th}_\mathrm{CP}(x_1;d)$ for $T=300$ K (solid blue line)
  and $T=600$ K (dotted red line) is plotted.}
\label{fig1}
\end{figure}

{\it Estimation of the Casimir-Polder force}.
Since in the presented scheme the information about $\delta$ is acquired by the 
continuous tunneling of atoms between the wells, 
the interferometer is suited for measuring forces which 
decay on a scale of typical inter-well distances of few microns. 
As an example, we analyze here the measurement of the Casimir-Polder force. 
We consider an experimentally realistic setup, 
consisting of a BEC of $N=2500$ $^{87}$Rb atoms trapped in a double-well potential, with its minima
separated by $l=4.8$ $\mu $m and the tunneling energy equal to $E_J/\hbar = 52.3\ s^{-1}$. 
When a sapphire plate is placed at a distance $d$ from one of the wells, 
a Casimir-Polder force acts on the atoms. The exact form of the potential, 
given by Eq.(26) of \cite{Antezza_2004},
depends on the dielectric properties of the atoms and the plate, as well as on the 
temperature $T$ of the latter.
If the thermal wavelength $\lambda_{th}=\frac{\hbar c}{k_BT}$ 
of the photons emitted from the plate is much larger than $d$
(as it is for $d\simeq5\ \mu$m when $T\leq100$ K), the 
Casimir-Polder potential is well approximated by 
$V_\mathrm{CP}(x_1;d)=-\frac{0.24\ \hbar\ c\ \alpha_0}{(x_1+l/2+d)^4}\ \frac{\varepsilon_0-1}{\varepsilon_0+1}$.
Here, $\varepsilon_0=9.4$ is the static value of the dielectric function of the sapphire, 
$c\simeq3\times10^8\ \frac{\mathrm{m}}{\mathrm{s}}$
is the speed of light and $\alpha_0=47.3\times10^{-30}$ m$^{3}$ 
is the static value of $^{87}$Rb atomic polarizability. 
The potential shifts the energy minima by $\delta$, calculated as in
\cite{nota_integrals} using $V(x_1)=V_\mathrm{CP}(x_1;d)$.
If the plate is positioned at $d=4\ \mu$m, then $\delta/\hbar=4.4\ s^{-1}$ and
$\frac{\delta}{E_J}=0.084$.
The period of Rabi oscillations is given by the detuned Rabi frequency and equals 120 ms. 

If the temperature of the plate is high, so that the condition 
$\lambda_{th}\gg d$ is no more satisfied, the Casimir-Polder 
interactions are described by a temperature-dependent potential
$V^{th}_\mathrm{CP}(x_1;d)=-\frac{k_BT\ \alpha_0}{4\ (x_1+l/2+d)^3}\ \frac{\varepsilon_0-1}{\varepsilon_0+1}$.
For $T\geq300$ K,  at distance $d$ of few microns, the thermal potential 
$V^{th}_\mathrm{CP}(x_1;d)$
vastly dominates over the low-temperature one, $V_\mathrm{CP}(x_1;d)$.

In Fig.\ref{fig1},  we plot the values of $\delta/\hbar$, as a function of distance $d$, calculated with
either $V_\mathrm{CP}(x_1;d)$ (dahsed black line) or
$V^{th}_\mathrm{CP}(x_1;d)$ for $T=300$ K (solid blue line) and $T=600$ K (dotted red line).
The error bars around the dashed line give the uncertainty $\Delta\delta_\mathrm{ ML}/\hbar$ of the
Rabi interferometer fed by a CSS
for a fit to $k=10$ points $t_i=\frac{2\pi i}{\omega k}$ in first Rabi period  with $m=10$ measurements at each time. 
In the calculation of the sensitivity, we alse include realistic experimental noise originating from two sources discussed
below.

{\it Sources of noise}.
Spin-polarized fermions are optimal candidates for the implementation of the above interferometric scheme
since the s-wave particle-particle interaction is naturally suppressed by the Pauli exclusion principle.
In the case of BEC, the atomic interactions can be reduced using the Feshbach resonance method.
This technique does not allow for tuning the two body interactions precisely
to zero. Some residual interactions always persist and
can spoil the sensitivity of the estimation. 
This can be taken into account by an additional term
$E_C\hat J_z^2$ in Hamiltonian (\ref{ham}). If the amplitude of the interactions
is small, $\gamma\equiv N\frac{E_C}{E_J}\ll1$, one can calculate the correction to the evolution
operator (\ref{evo_new}) in first order Dyson expansion with
respect to small parameter $\gamma$. 
Using realistic value $\gamma=0.1$ \cite{nota_inter}, 
the contribution to the error bars in Fig.\ref{fig1} is negligible.

An important source of noise in the Rabi interferometer is given by the limited resolution of
population imbalance measurement. 
This can be taken into account by substituting the ideal probability $p(\{n\}|\delta)$
with the convolution $p_{res}(\{n\}|\delta) = \sum_{n'} \mathcal{P}(n, n') p(\{n'\}|\delta)$, where 
$\mathcal{P}(n, n')$ gives the probability to measure the population imbalance $n'$, given the true value $n$. 
As an example, we take 
$\mathcal{P}(n, n') = \frac{1}{\sqrt{2\pi}\sigma_{\mathrm{res}}}\exp{\big[-\frac{(n-n')^2}{2\sigma_{\mathrm{res}}^2}\big]}$, 
with a conservative value $\sigma_{\mathrm{res}}=40$ 
(the population imbalance is measured with a resolution of $\pm40$ particles). 
Then, the ratio $\frac{\Delta\delta_{\mathrm{ML}}}{\delta}$ 
for the spin coherent state with 2500 atoms increases by a factor of two and this value of $\sigma_{\mathrm{res}}$
is used to present the sensitivity in Fig.\ref{fig1}.
Even in presence of this noise, the sensitivity is sufficient to precisely distinguish between 
thermal and zero-temperature regimes of the Casimir-Polder force. This is the main result of this Letter. 

\begin{figure}[htb!]
 \includegraphics[clip, scale=0.37]{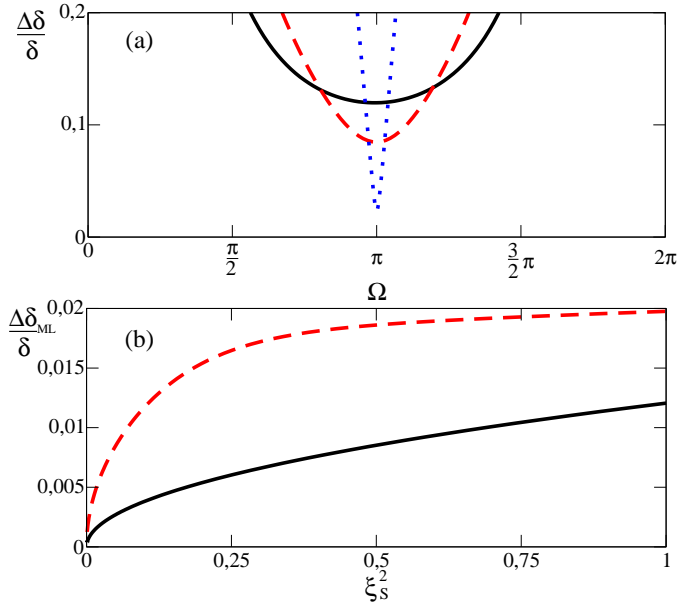}
 \caption{(color online): (a) The relative uncertainty $\frac{\Delta\delta}{\delta}$
   as a function of $\Omega$ for $m=1$. Solid black, dashed red and 
   dotted blue lines correspond to $\xi^2=$1.0, 0.5, 0.017, respectively. For $\Omega\simeq\pi$, 
   the sensitivities reach optimal point.
   (b) The relative uncertainty $\frac{\Delta\delta_{\rm ML}}{\delta}$ as a function $\xi^2$. 
   The solid black line corresponds to 100 measurements at optimal point. 
   The dashed red line is a result of a fit
   to $k=$10 different points with $m=$10 repetitions at each point.
   Here, $N = 2500$ and $\frac{\delta}{E_J}=0.084$.}
 \label{fig2}
\end{figure}

{\it Interferometer with squeezed input states}.
So far, we have discussed the sensitivity of the Rabi interferometer fed with a coherent input state.
Below we show that higher sensitivity can be reached when using
spin squeezed states. 
Such states can be created by adiabatically splitting an interacting BEC trapped in a 
double well potential, as recently experimentally demonstrated in \cite{Esteve_2008}.   

When $N$ is large, and if the input states $|\psi_{inp}\rangle = \sum_{n=0}^N c_n |n, N-n \rangle$
are symmetrical, i.e. $c_n=c_{N-n}$, one can approximate $c_n$'s with Gaussians, 
$c_n\propto e^{-(n-\frac N 2)^2/(4\sigma^2)}$. Direct calculation gives
$\langle\hat J_x\rangle\simeq\frac{N}{2}e^{-\frac{1}{8\sigma^2}}$, 
$\langle\hat J^2_{x,y}\rangle\simeq\frac{N^2}{8}(1\pm e^{-\frac{1}{2\sigma^2}})$ and 
$\langle\hat J^2_{z}\rangle\simeq\sigma^2$.
According to the definition of spin squeezing \cite{Wineland_1994}, 
$\xi^2=N\frac{\langle \hat J_z^2\rangle}{\langle \hat J_x\rangle^2} 
\approx 4 \frac{\sigma^2}{N}e^{\frac{1}{4\sigma^2}}$, 
states with $\sigma<\frac{\sqrt N}{2}$ are called squeezed states and the values of $\xi^2$ range from 1 for a 
CSS, to 0 for a Fock state. 

In Fig.\ref{fig2}(a) we plot the sensitivity from Eq.(\ref{err}) for a single measurement ($m=1$) 
in units of $\delta$ for three values
of squeezing, as a function of time in the first Rabi period. We use realistic parameters
$N = 2500$ and $\frac{\delta}{E_J}=0.084$ from previous section.
Squeezing of the input state clearly improves the precision. 

We also observe that the interferometer
has an optimal working time equal to $\frac\pi\omega$. This suggests the following
strategy for estimation of $\delta$. Instead of fitting a curve to a set of equally distributed points in time,
one should focus the experimental effort around the working point. 
In Fig.\ref{fig2}(b)
we compare the sensitivities $\frac{\Delta\delta_{\rm ML}}{\delta}$
as a function of squeezing parameter $\xi^2$ for these two estimation strategies. The solid black line is
obtained using the optimal strategy while the dashed red line results from a fit
to $k=10$ points equally spaced in first Rabi period. 
Even though the optimal time strategy proves better, the difference is smaller than
one order of magnitude. Notice that for both protocols precision of the
estimation strongly benefits from squeezing of
the input state. 
For the optimal time strategy with a generic symmetrical input state, this can be shown analytically by expressing
the sensitivity in terms of the spin squeezing parameter 
$\frac{\Delta\delta_{\rm ML}}{\delta}=\xi\frac{\omega}{2\delta\sqrt N\sqrt{mk}}$. 

By finding minimum of $\Delta\delta_{\mathrm{ML}}$ with respect to $\sigma$, one can also show that for a Fock
state the sensitivity reaches Heisenberg limit, i.e.  $\Delta\delta_{\rm ML}\propto N^{-1}$.
Recently, a squeezed state with $\xi^2=0.58$ for $N\simeq2000$ particles was
achieved \cite{Esteve_2008}. As seen in Fig.\ref{fig2}(b), such state would allow for
1.5 times improvement of sensitivity in case of optimal point
estimation strategy. 

{\it Comparison with other interferometric setups}.
The possibility to use cold/degenerate atoms for the measurement of forces at small distances  
has lead to a number of proposals and experiments 
\cite{Carusotto_2005, Sorrentino_2009, Ferrari_2006, Wolf_2007, Pasquini_2004, Harber2005, Lin_2004, Obrecht_2007}.  
In \cite{Obrecht_2007}, the gradient of the Casimir-Polder force was deduced from the shift of the 
frequency of the collective oscillations of a BEC in a trap put below a surface. 
The precision of the experiment was not sufficient to make a distinction
between thermal and zero-temperature regimes of the force. On the theoretical side, 
the References \cite{Carusotto_2005, Sorrentino_2009}, 
propose to estimate the strength of the interaction between the atoms and a surface using the 
frequency shift of Bloch oscillations of a cold fermionic gas in a vertical optical lattice. 
An important aspect of this proposal is the scaling of the sensitivity 
$\Delta \delta \sim t^{-1/2}$ with the oscillation time $t$.
As argued before, this is not the case of the Rabi interferometer. However,
when compared to the Bloch oscillations proposal, our setup has the advantage of
the sensitivity scaling with the number of particles in the input state $\Delta \delta \sim N^{-\beta}$.
As shown above, this scaling is at the shot noise $\beta=1/2$ for a CSS 
and can be further increased $1/2<\beta\leq1$ for entangled atoms. 

{\it Conclusions}.
We have shown that a degenerate either bosonic or fermionic gas in a double-well potential
can realize a sensitive device for measuring short-range interactions, such as the Casimir-Polder force. 
Even the interferometer is fed with a classical spin coherent state with moderate number of atoms, and in the presence of
realistic experimental noise, the Casimir-Polder force can be measured
with precision sufficient to distinguish between its thermal and zero-temperature regimes. 
The estimation protocol further benefits from squeezing the input state and measuring the population imbalance at the
optimal time.


\begin{thebibliography}{27}
\bibitem[{not({\natexlab{a}})}]{nota_integrals}
\bibinfo{note}{We have $E_J= - 2 \int \ud x_1 \, \psi_{a}(x_1) [
  -\frac{\hbar^2}{2m} \frac{\ud^2}{\ud ^2 x_1} +V_{dw}(x_1) ] \psi_{b}(x_1)$,
  where $V_{dw}(x_1)$ is the double well potential and $m$ the mass of the
  atoms. The parameter $\delta = \int \ud x_1 \, \big( |\psi_{a}(x_1)|^2 -
  |\psi_{b}(x_1)|^2 \big) V(x_1)$, where $V(x_1)$ is the perturbing potential.}
\bibitem[{not({\natexlab{d}})}]{nota_Fisher}
\bibinfo{note}{The Fisher information is defined as $\mathcal{F}=\int
  dn_1\ldots dn_k\
  \frac{1}{p(\{n\}|\delta)}\left[\frac{\partial}{\partial\delta}p(\{n\}|\delta%
)\right]^2.$}.
\end{thebibliography}

\end{document}